\documentclass[prd,floatfix,preprintnumbers]{revtex4}
\usepackage{amssymb}
\usepackage{amsmath}
\usepackage{subfigure}
\usepackage{graphicx}
\usepackage{xcolor}

\begin{document}

\title{Big bang nucleosynthesis with rapidly varying G}

\author{Anish Giri}
\email{anish.giri@vanderbilt.edu}
\affiliation{Department of Physics and Astronomy, Vanderbilt
  University, Nashville, TN, 37235, USA}
\author{Robert J. Scherrer}
\email{robert.scherrer@vanderbilt.edu}
\affiliation{Department of Physics and Astronomy, Vanderbilt
  University, Nashville, TN, 37235, USA}
\date{\today }

\begin{abstract}
We examine big bang nucleosynthesis (BBN) in models with a time-varying gravitational constant $G$, when
this time variation is rapid on the scale of the expansion rate $H$, i.e, $\dot G/G \gg H$.  Such
models can arise naturally in the context of scalar-tensor theories of gravity and result in additional
terms in the Friedman equation.  We examine two representative models:  a step-function evolution
for $G$ and a rapidly-oscillating $G$.  In the former case, the additional terms in the Friedman equation
tend to cancel the effects of an initial value of $G$ that differs from the present-day value.
In the case of deuterium, this effect is large enough to reverse the sign of the change in (D/H) for a given
change in the initial value of $G$.  For rapidly-oscillating $G$, the effect on the Friedman equation
is similar to that of adding a vacuum energy density, and BBN allows upper limits to be placed
on the product of the oscillation frequency and amplitude.  The possibility that
a rapidly oscillating $G$ could mimic a cosmological constant is briefly discussed. 
\end{abstract}

\maketitle

\section{Introduction}

The possibility that the gravitational constant $G$ might evolve with time has long been a topic of
interest, dating back to Dirac's large numbers hypothesis \cite{Dirac}.  Many limits
can be placed on the time variation of $G$, including constraints from the cosmic microwave background
\cite{Ooba}, binary pulsars \cite{Zhu}, seismology of the sun \cite{Guenther} and other stars \cite{Bellinger},
and precision measurements of the orbits of Mars \cite{marslimit} and the Moon \cite{moonlimit}.  (For reviews,
see Refs. \cite{Uzan,Will}).  While the solar system measurements \cite{marslimit,moonlimit} give the tightest limits on $\dot G/G$,
they can constrain this variation only over the course of the past few decades.  The physical system that
provides limits on the time variation of $G$ over the longest timeline is big bang nucleosynthesis (BBN), which is sensitive to
changes in $G$ between the first few seconds of the Universe and the present day.

BBN limits on the time variation of $G$ have been widely investigated
\cite{Ser96,San97,Dam99,Che01,Krauss,Pet04,Bambi,Clifton,Coc,Alvey,Yeh}.  Nearly all of these papers
assume a slowly varying value for $G$, in the sense that $\dot G/G \ll H$, where $H$ is the Hubble parameter, or
they simply take $G$ to be constant during BBN, with a value different from its present-day value.
An exception is the discussion of Accetta and Steinhardt \cite{Accetta}, who examined the cosmological
effects of a rapidly oscillating $G$.  (See also Ref. \cite{W-S}). Our study represents a reexamination and extension of this early work.
We investigate two variations on this idea proposed in the literature, a step-function variation
in $G$ and a rapidly oscillating $G$, and we demonstrate their effect on BBN.

As is well known, time variation of fundamental constants is only well defined when the constants in question
are dimensionless, a fact which is often glossed over in discussions of the time variation of $G$.  A
variation in $G$ corresponds to a change in the Plank mass $M_{Pl} = 1/\sqrt{G}$, so that a time variation in
$G$ can be modeled as a change in the ratio of $M_{Pl}$ to all of the other masses and energy scales in the Standard
Model.  This point is discussed in more detail in Ref. \cite{Alvey}.

In the next section, we discuss our particular models for rapidly-varying $G$.  In Sec. III, we calculate the changes in
primordial element production, highlighting the effects unique to rapid
time variation of $G$.  In Sec. IV we discuss our main results.  We find that the  modifications
to the Friedman equation that are unique to rapidly-varying $G$ have a profound effect on the predicted primordial
element abundances.  For a step-function variation in $G$, the additional terms in the Friedman equation tend
to partially cancel the effect on BBN of an initially-different value of $G$, while for rapidly-oscillating $G$ we
can place upper limits on the product of the oscillation frequency and amplitude.

\section{Models with Rapidly-Varying G}

In the standard zero-curvature cosmological model, the evolution of the scale factor $a$ is given by
the Friedman equation:
\begin{equation}
\label{Friedman}
\left(\frac{\dot a}{a}\right)^2 = \frac{8 \pi G \rho}{3},
\end{equation}
where $\rho$ is the total energy density in the Universe, and the overdot denotes derivative with
respect to time.
An adiabatic change in $G$, for which $\dot G/G \ll H \equiv \dot a/a$ can be modeled by simply
taking $G$ to be the appropriate function of time in Eq. (\ref{Friedman}).  However, this approach is
inadequate when $\dot G/G \gg H$.
In Ref. \cite{Accetta}, Accetta and Steinhardt present a variant of the Friedman equation for the case of
rapidly-varying $G$, namely
\begin{equation}
\label{A-S}
\left(\frac{\dot a}{a}\right)^2 = \frac{\dot G}{G} \frac{\dot a}{a} + \frac{8 \pi G}{3} (\rho_M + \rho_\phi),
\end{equation}
where $\rho_M$ is the standard background energy density (matter plus radiation),
and $\rho_\phi$ is the contribution from a scalar field $\phi$ that couples to gravity and provides the change in $G$.
The prototypical models that produce this type of evolution are scalar-tensor theories of gravity, with
action (see, e.g., Ref. \cite{extended})
\begin{equation}
S= \int d^4 x \sqrt{-g}\left[\frac{1}{2}F(\phi)R - \frac{1}{2} \omega(\phi) \phi^{;\mu} \phi_{;\mu} - V(\phi)
+ L_M\right].
\end{equation}
Here $R$ is the Ricci scalar, $V(\phi)$ is the scalar field potential, and $L_M$ is the Lagrangian for the
matter.  Taking $\omega(\phi) = 1$, the value of $G$ is given in terms of $F$
by
\begin{equation}
\frac{1}{8 \pi G} = F(\phi),
\end{equation}
and the Friedman equation reduces to Eq. (\ref{A-S}),
with the scalar field density taking the usual
form
\begin{equation}
\rho_\phi = \frac{\dot \phi^2}{2} + V(\phi).
\end{equation}
Following Ref. \cite{Accetta}, we will consider only the case where $\rho_\phi \ll \rho_M$, with the result
that Eq. (\ref{A-S}) can be solved for $\dot a/a$:
\begin{equation}
\label{maineq}
\frac{\dot a}{a} = \frac{1}{2} \frac{\dot G}{G} + \left[\frac{8 \pi G}{3}\rho_M + \frac{1}{4}
\left(\frac{\dot G}{G}\right)^2 \right]^{1/2}.
\end{equation}
We will assume that $G$ evolves to its present-day value, $G_N$, with $\dot G = 0$,
shortly after BBN, so our models can be assumed to satisfy all of the non-BBN constraints on the time-variation of $G$.

In what follows, we will consider two representative models for the time evolution of $G$.  In the first
model, we assume that $G$ evolves approximately as a step function, with initial
value $(1+A) G_N$ for $t < t_i$,
where A is a dimensionless constant, 
and $G = G_N$ at $t > t_f$, with linear evolution between these two values.  Then the full expression for $G(t)$
is
\begin{eqnarray}
\label{step1}
G/G_N &=& (1+A),~~~~~~~~~~~~~~~~~~~~t < t_i, \\
\label{step2}
G/G_N &=& 1 + A \left(\frac{t_f - t}{t_f - t_i}\right), ~~~~~~~t_i < t < t_f, \\
\label{step3}
G/G_N &=& 1,~~~~~~~~~~~~~~~~~~~~~~~~~~~~t > t_f.
\end{eqnarray}
Step function models for $G$ have been explored
in connection with inflation \cite{Ash} and, more recently,
as a possible resolution of the Hubble tension \cite{Marra,Sakr,Benevento,Kable,Alestas1,Alestas2}.
Other observational consequences of a step function change
in $G$ are explored in Refs. \cite{Alestas1,Alestas2,Paraskevas}.
Models of this kind are easily generated in the context of
scalar-tensor theories with a
step function $F(\phi)$ (see, e.g., Refs. \cite{Ash,Paraskevas}),
and they naturally asymptote to $G = G_N$, with $\dot G = 0$, at late times.
In general, the value of $A$ is taken to be a free parameter in these models, but
models that resolve the Hubble tension typically require a change in $G$ on the order
of 10\%.

A second set of models we examine are those with rapidly oscillating G,
which we take to have the form
\begin{equation}
\label{oscillate}
G/G_N = 1 + A \cos(\omega t + \phi),
\end{equation}
where $A$, $\phi$, and $\omega$ are constants. Such models arise naturally
in scalar-tensor theories when the scalar field oscillates in the minimum of
a potential.  This was the model
originally discussed in Ref. \cite{Accetta}, and these models were also examined
in the context of dark energy in Ref. \cite{LiScherrer}.
One can construct more general models in which $\omega$ evolves with time, but for simplicity,
and to pinpoint how oscillating models affect BBN, we will consider only the case
where $\omega$ can be treated as constant during the epoch of BBN.
Unlike the case
of a step function, we cannot take Eq. (\ref{oscillate}) to apply all of the way up to the present day.
While the average value of $G$ is always equal to $G_N$, Eq. (\ref{maineq}) shows
that any oscillation that is ``rapid" at the epoch of BBN ($\dot G/G \sim {\rm s}^{-1}$)
will dominate the expansion at late times,
producing an expansion law completely at odds with observation.  Hence, we assume that Eq. (\ref{oscillate})
applies only at early times, with $G$ evolving toward a constant value of $G_N$ at some time after BBN.

\section{Effects on Big Bang Nucleosynthesis}

In standard BBN, element production takes place in two stages (see, e.g., Refs. \cite{Yeh,FOYY}
for recent discussions).
At early times, 
($T \gtrsim 1$ MeV) the weak interactions
maintain a thermal equilibrium ratio between neutrons and protons: 
\begin{eqnarray}
n~+~\nu _{e} &~\longleftrightarrow ~&p~+~e^{-},  \notag
\label{weak-interactions} \\
n~+~e^{+} &~\longleftrightarrow ~&p~+~\overline{\nu }_{e},  \notag \\
n &~\longleftrightarrow ~&p+e^{-}~+~\overline{\nu }_{e},
\end{eqnarray}
while a thermal abundance of deuterium is maintained by 
\begin{equation}
n~+~p~\longleftrightarrow ~\mathrm{D}~+~\gamma .
\end{equation}
The weak reactions drop out of thermal equilibrium at $T\sim 1$ MeV, while
free neutron decay continues until $T\sim 0.1$ MeV, when rapid fusion into
heavier elements occurs. At this point, nearly all of the remaining neutrons become bound
into $^{4}$He, with a small fraction incorporated into
deuterium, tritium, and $^3$He.  There is also production of $^{7}$Li and $^{7}$Be, with the latter
decaying into the former via electron capture at the beginning of the
recombination era \cite{Sunyaev}.
The element abundances produced in BBN
depend on the baryon/photon ratio $\eta $, which can be independently
determined from the CMB. We adopt a value of $\eta = 6.1 \times 10^{-10}$,
consistent with results from Planck \cite{Ade}. This value of $\eta$
yields predicted abundances of D and $^{4}$He consistent with observations.
On the other hand, the $^7$Li abundance predicted by BBN is well known to lie
a factor $\sim 3$ above the
observationally-inferred value, giving rise to the so-called ``lithium problem" \cite{lithium}.
As it is not clear whether this problem lies in the interpretation of the lithium abundance observations or
in the standard model of BBN itself,
we will concentrate on $^4$He and deuterium production.  Following
standard practice, we give the primordial $^4$He abundance 
as a mass fraction, $Y_p$, and the  deuterium abundance as a number density relative
to hydrogen, (D/H).

When the change in $G$ is taken to be a constant during BBN, the effect on the $^4$He and deuterium abundances
is well fit by \cite{FOYY}
\begin{equation}
\label{YconstG}
Y_p = Y_{p0} (G/G_N)^{0.357},
\end{equation}
and
\begin{equation}
\label{DconstG}
{\rm (D/H)} = {\rm (D/H)}_{0} (G/G_N)^{0.952},
\end{equation}
where the $0$ subscript denotes the standard values of these abundances with an unmodified gravitational constant.
Although both the $^4$He and deuterium abundances are increasing functions of $G$, this dependence arises for
different reasons in these two cases.
The $^4$He abundance depends mostly on the abundance of free neutrons
when rapid fusion commences.  An increase in $G$ causes the weak rates to freeze out at
a higher temperature, when the equilibrium abundance of free neutrons is larger.  Since nearly all of these
neutrons end up in $^4$He, the result is a larger $^4$He abundance.  However, changing $G$ also has a second
effect on the $^4$He abundance.  A larger $G$ corresponds to a smaller Hubble time at a given temperature.
Thus, the temperature at which rapid fusion begins corresponds to a smaller time
over which free neutron decay can occur, giving more neutrons to bind into $^4$He.  This dual dependence
on $G$ can be seen in Ref. \cite{Bambi}, which presents response functions for the various element abundances
as a function of a change in $G$ that is effectively a delta function of time.  For $Y_p$, the response function is always
positive, with dual peaks at $T \sim$ 1 MeV and $T \sim 0.1$ MeV, corresponding to the two effects noted above.
Deuterium, on the other hand, is sensitive almost entirely to the value of $G$ when fusion begins.
Given an arbitrarily long time, essentially all of the deuterium would fuse into heavier nuclei; the
existence of primordial deuterium is a result of the relatively short time over which fusion can occur.  Increasing
$G$ decreases this time, resulting in less fusion of deuterium into heavier elements and a corresponding
increase in (D/H).  This is evident in the response function for deuterium given in Ref. \cite{Bambi}.
This response function is positive and sharply peaked near $T \sim 0.1$ MeV.

\begin{figure}[!ht]
	\begin{subfigure}
	\centering
	\includegraphics[width=3in]{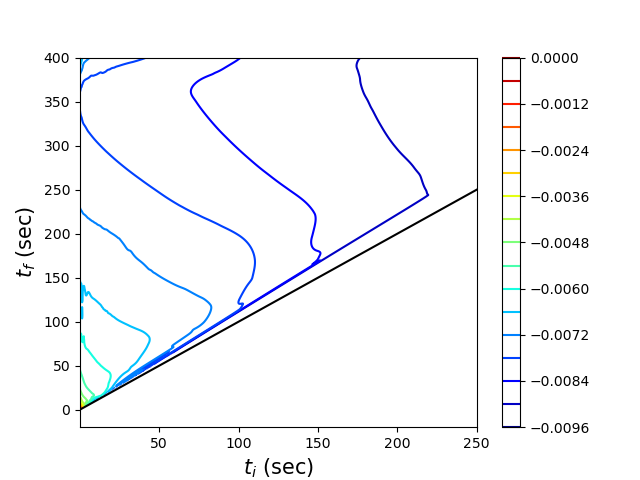}
	\end{subfigure}
	\begin{subfigure}
	\centering
	\includegraphics[width=3in]{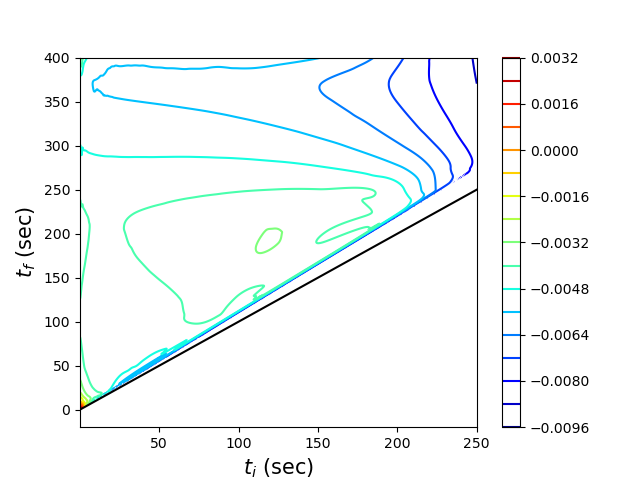}
	\end{subfigure}
	\caption{Change in the $^4$He mass fraction, $\Delta Y_p$, for a step-function variation in $G$
(Eqs. \ref{step1}$-$\ref{step3}) as a function of $t_i$ and $t_f$, with $A = -0.1$
using (left) only
the standard Friedman equation (\ref{Friedman}) and (right) the full Friedman equation (\ref{maineq}) including the appropriate $\dot G/G$ terms.}
\end{figure}

\begin{figure}[!ht]
	\begin{subfigure}
	\centering
	\includegraphics[width=3in]{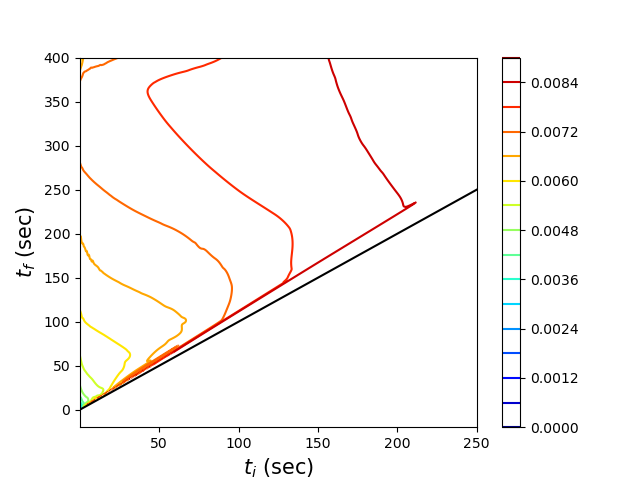}
	\end{subfigure}
	\begin{subfigure}
	\centering
	\includegraphics[width=3in]{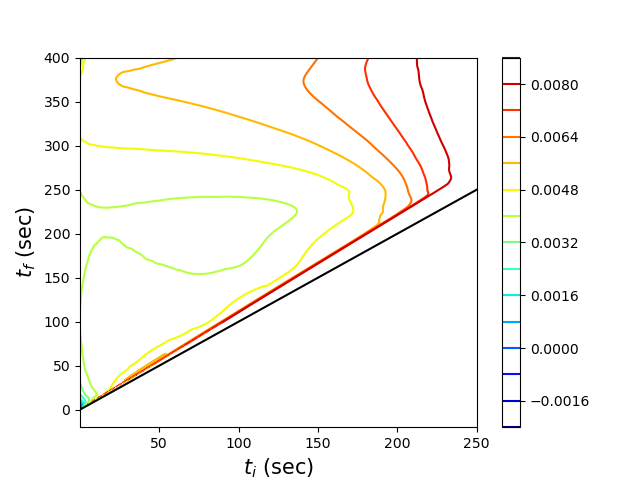}
	\end{subfigure}
	\caption{As Fig. 1, with $A = +0.1$.}
\end{figure}

To calculate the primordial element abundances, we have modified PRyMordial, a Python code for BBN calculations
\cite{primordial}, incorporating Eq. (\ref{maineq}) for the Friedman equation.  To highlight the
specific effects of rapid changes in $G$, we have also calculated the primordial element abundances using
the standard Friedman equation (\ref{Friedman}) with time-varying $G$.
We have chosen to present our results in terms of the change in the primordial $^4$He mass fraction,
$\Delta Y_p$, and the change in the number density of deuterium relative to hydrogen, $\Delta$(D/H), due
to the time variation in $G$, rather than
giving the altered values of $Y_p$ and (D/H) themselves.  This approach has the advantage that
$\Delta Y_p$ and $\Delta$(D/H) are less sensitive to the input parameters such as the baryon-to-photon ratio, the neutron
lifetime, and changes to the nuclear cross sections,
because altering these parameters will tend to change both the standard-model and nonstandard values of these abundances in roughly
the same way. Although we provide constraints on $\dot G/G$ for the rapidly-oscillating models, such constraints are subject
to future changes in the observed element abundances, so we consider our results for $\Delta Y_p$ and $\Delta$(D/H) to be more interesting
and fundamental.

Consider first the step-function evolution for $G$ (Eqs. \ref{step1}$ - $\ref{step3}).
As it is impractical to scan over the entire parameter space, and we are most interested
in the effect of including the $\dot G/G$ terms in the modified Friedman equation,
we consider only two representative values for $A$ ($-0.1$ and $+0.1$) and scan over $t_i$
and $t_f$.
Our results for the $^4$He
abundances in this case with $A = -0.1$ and $A = +0.1$ are displayed in Figs. $1-2$.  First consider the effect of simply altering
$G$ in the standard Friedman equation without including the $\dot G/G$ terms.  When $t_i$ is large, this is simply equivalent
to BBN with a different (constant) value of $G$: here $G = (0.9)(G_N)$ (Fig. 1, left) or $G = (1.1)(G_N)$ (Fig. 2, left).
As we have
already noted, $^4$He is most sensitive to the value of $G$ at early times; for large $t_i$, we simply obtain a constant
increase or decrease in the $^4$He abundance, given by the corresponding constant change in $G$ as in Eq. (\ref{YconstG}).
However, when we use the full
expression for the scale factor given by Eq. (\ref{maineq}) (Figs. 1-2, right), we see that the $\dot G/G$ terms strongly modify the change in
the primordial $^4$He production, producing more complex behavior.  In particular, for
$A < 0$, we have $\dot G/G > 0$, so the $\dot G/G$ terms increase the value of $\dot a/a$.  This partially cancels the effect
of the smaller initial value of $G$, producing a smaller decrease in $Y_p$.  The
opposite effect occurs when $A = +0.1$. In this case, $\dot G/G$ is negative, resulting
in a decrease in $\dot a/a$ that partially cancels the effect of the initially
larger value of $G$.  Thus, in both cases, the effect of including the additional $\dot G/G$ terms in the expression
for $\dot a/a$ in Eq. {(\ref{maineq}) is to reduce the overall effect of the change in $G$.

\begin{figure}[!ht]
	\begin{subfigure}
	\centering
	\includegraphics[width=3in]{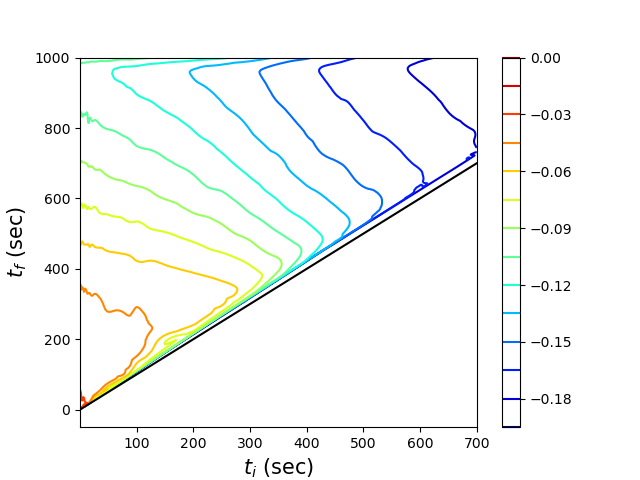}
	\end{subfigure}
	\begin{subfigure}
	\centering
	\includegraphics[width=3in]{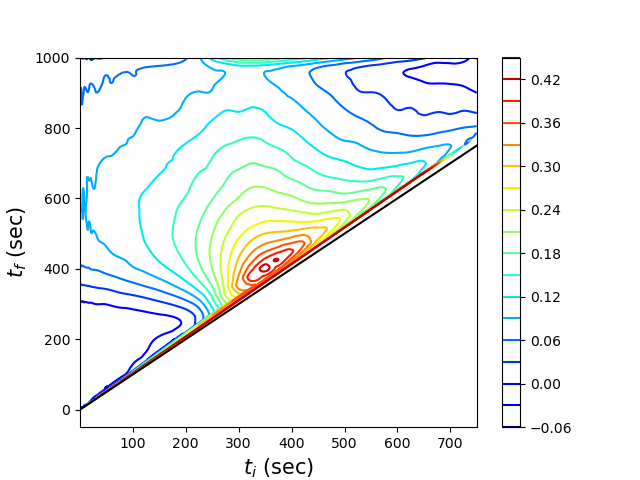}
	\end{subfigure}
	\caption{Change in D/H expressed as $\Delta$(D/H)$\times 10^5$ for a step-function variation in $G$
(Eqs. \ref{step1}$ - $\ref{step3}) as a function of $t_i$ and $t_f$, with $A = -0.1$
using (left) only
the standard Friedman equation (\ref{Friedman}) and (right) the full Friedman equation (\ref{maineq}) including the appropriate $\dot G/G$ terms.}
\end{figure}

\begin{figure}[!ht]
	\begin{subfigure}
	\centering
	\includegraphics[width=3in]{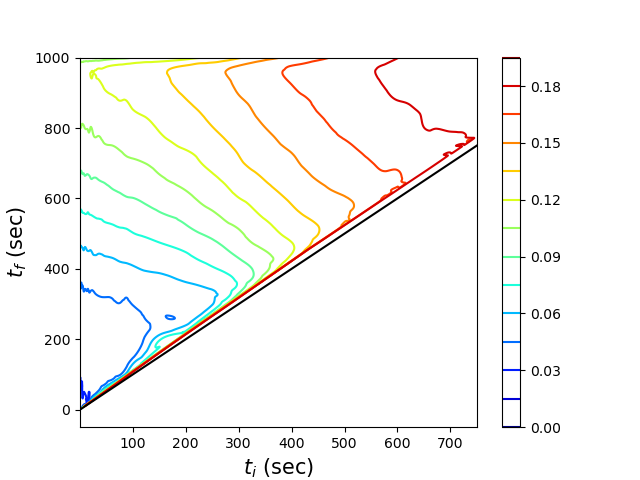}
	\end{subfigure}
	\begin{subfigure}
	\centering
	\includegraphics[width=3in]{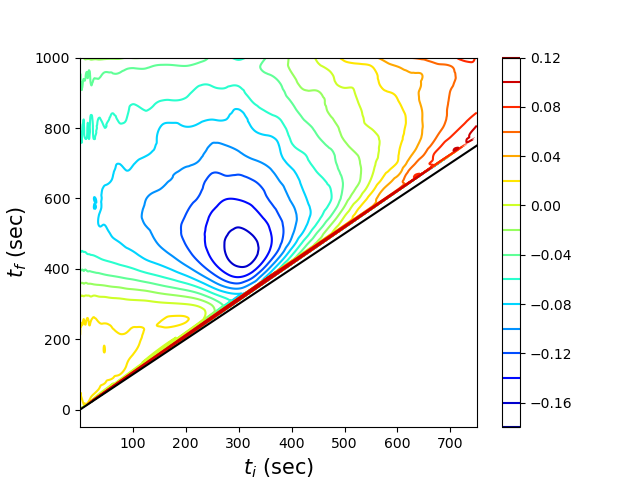}
	\end{subfigure}
	\caption{As Fig. 3, for $A = +0.1$.}
\end{figure}

Turning to deuterium (Figs. $3-4$), we see a similar effect but with an even
larger magnitude.  As expected, a simple step function change in $G$ in
the unmodified Friedman equation results in change in (D/H) with the same
sign as the change in $G$.  However, making use of Eq. (\ref{maineq})
gives an additional contribution to $\dot a/a$ with the opposite sign, as it does
for $^4$He.  In the case of deuterium, this effect is so large that
it reverses the sign of the change in (D/H) over almost all of the parameter
space:  an initial value of $G$ larger than $G_N$ gives a reduction in (D/H), while an
initial $G$ smaller than $G_N$ corresponds to an increase in (D/H).

One might naively assume that the effect of adding these $\dot G/G$ terms
to the Friedman equation would be maximized for the smallest possible
values of $t_f - t_i$, for which the change in $G$ most closely approximates
a step function and $\dot G/G$ is maximized. While decreasing $t_f - t_i$
does increase $\dot G/G$, it also reduces
the range in time over which $\dot G/G$ is nonzero.  There is a trade-off
between these two effects, which
produces a much more complex dependence of the element abundances
on $t_i$ and $t_f$ in Figs. $1-4$ than does a simple change in $G$ in Eq. (\ref{Friedman}).

While we have concentrated on the deuterium and $^4$He abundances, we have also checked to see whether there
are any parameter values that significantly reduce the predicted $^7$Li abundance while at the same time
giving acceptable abundances for deuterium and $^4$He, thus serving as a
possible solution to the lithium problem.  We find no such parameter values for the step function change in
$G$.  In general, a significant reduction in $^7$Li is always accompanied by an unacceptably large
increase in the predicted deuterium abundance. A similar effect has been seen in other proposed
solutions to the lithium problem using nonstandard BBN (see, e.g., Ref. \cite{lithium}).

A full set of constraints on the allowed parameter space would require a scan over the value
of $A$ as well as $t_i$ and $t_f$ and would, in any case, not provide much insight into this model.  However,
we can use our limited set of results to make useful qualitative statements about BBN constraints
on this model.  In going from a constant change in $G$ to a step-function change without
including the $\dot G/G$ terms in the Friedman equation, it is clear from Figs. $1-2$
that the step function always produces a smaller change in $^4$He and deuterium than does the same
constant change in $G$.  This is to be expected, since the only effect of the step function in this case
is to reduce the time over which $G$ differs from its present-day value.
Thus, any observational limits for a particular value of $A$ will
always be weaker in the step-function case than in the case where the change in $G$
is constant throughout nucleosynthesis.  In going from the naive step-function model to one which correctly
incorporates the additional $\dot G/G$ terms in the Friedman equation, these limits are weakened even further
when considering $^4$He, since the effect of adding these additional terms tends to counteract the
effect of changing the value of $G$, as we have noted.  However, in the case of deuterium, the situation is
more complex.  In this case, the sign of the effect on the deuterium abundance can change, with the most extreme
values for $\Delta$(D/H) corresponding to $t_i = 200-500$ sec and $t_f = 400 - 700$ sec.  Over this range, observational
limits on deuterium have the potential to place tighter constraints on a step-function variation
in $G$ than on models with a constant $G$.  However, this range for $t_i$ and $t_f$ represents
only a small slice of parameter space for these models.

Now consider the effect of an oscillatory change in $G$ (Eq. \ref{oscillate}).
We can distinguish three regimes.  For $\omega^{-1} > 10^3$ s, the value of $G$
is roughly constant during BBN with $\dot G/G < H$ throughout BBN, so Eqs. (\ref{YconstG}) and (\ref{DconstG}) give
the primordial $^4$He and deuterium abundances in this case,
with well-known limits already derived in the literature.
For $\omega^{-1} \sim 1 - 10^3$ s, we find that the element abundances
are a complicated function of $A$, $\omega$, and $\phi$ in Eq. (\ref{oscillate}), with
results that do not provide much insight.  The
case $\omega^{-1} < 1$ s is more amenable to the derivation of
limits on the model parameters.  In this
case, the first term in Eq. (\ref{maineq}) averages to zero, and the effective
density is simply increased by an amount proportional to $(\dot G/G)^2$.
This density increase is constant (or at least can be taken to be constant during BBN; it must, of course
decay away at some time after BBN).
Expressing this effective change in density in terms of a constant $\rho_0$, we have
\begin{eqnarray}
G\rho_0 &=& \frac{3}{32 \pi} \left(\frac{\dot G}{G}\right)^2,\\
\label{Aomega}
&=& \frac{3}{64 \pi} A^2 \omega^2,
\end{eqnarray}
where the second equality is valid for $A \ll 1$ and $\omega t \gg 1$, for Hubble time $t$.
It is clear that the effects of an oscillating $G$ with constant oscillation frequency $\omega$
cannot be modeled using the conventional parametrization in terms of an additional
effective number of relativistic degrees of freedom. Instead, the ratio
of $\rho_0$ to the background radiation density will be a sharply increasing function of time.
Hence, values of $(\dot G/G)^2$ large enough to be ruled out by the deuterium abundance
will have a negligible impact on $Y_p$, so we need consider only the former in deriving limits
on this model. (Conversely, an additional stiff component with $\rho \propto a^{-6}$ has a much
larger effect effect on the $^4$He abundance than on the deuterium abundance, so that
only limits on the former need be considered in that case \cite{stiff}). 

In the regime of interest for $\rho_0$, our numerical simulations are well fit by a change in (D/H) given by
\begin{equation}
\Delta {\rm (D/H)} = 1.9 (G \rho_0/{\rm s}^{-2})^{0.7},
\end{equation}
which can be expressed in terms of the parameters in Eq. (\ref{oscillate}) as
\begin{equation}
\label{Bomega}
\Delta {\rm (D/H)} = 0.10 (A^2 \omega^2 /{\rm s}^{-2})^{0.7}.
\end{equation}
Recent numerical simulations give, as the prediction of standard BBN (SBBN) \cite{YOF,Yeh},
\begin{equation}
{\rm (D/H)}_{\rm SBBN} = (2.51 \pm 0.11) \times 10^{-5},
\end{equation}
while observations yield \cite{YOF,Yeh}
\begin{equation}
{\rm (D/H)}_{\rm obs} = (2.55 \pm 0.03) \times 10^{-5}.
\end{equation}
Taken together, these suggest a very conservative upper bound on $\Delta {\rm (D/H)}$:
\begin{equation}
\Delta {\rm (D/H)} < 0.2 \times 10^{-5}.
\end{equation}
Then Eq. (\ref{Bomega}) gives an upper bound on the parameters in a model with rapidly
oscillating $G$ with constant oscillation frequency, namely
\begin{equation}
A \omega < 4.4 \times 10^{-4} ~{\rm s}^{-1},
\end{equation}
subject to the range over which our discussion is valid: $A \ll 1$ and $\omega^{-1} < 1$ s.

\section{Discussion}
While our investigation of the effects of a rapidly-changing $G$ on BBN is far from exhaustive,
it does provide some interesting new results.  For models motivated by scalar-tensor theories, the
effects of including the additional $\dot G/G$ terms in the Friedman equation are profound.
In the case of a step-function potential, these terms tend to cancel the effects of an initial
value of $G$ that differs from the present-day $G_N$; in the case of deuterium, we find that they can
even reverse the sign of the change in (D/H).  For a rapidly oscillating $G$ with constant
oscillation frequency, the effect is similar to adding a constant vacuum energy density.  This
dominantly effects deuterium production, allowing us to place an upper limit on the product
of the oscillation frequency and amplitude.

Our main results show that models with rapidly-varying $G$ provide very distinct cosmological signatures beyond
previously-considered models with adiabatic variations in $G$.  It would be interesting to examine these
effects during other cosmological epochs.  For example, it is tempting to consider whether a rapidly-oscillating $G$ with constant oscillation frequency could
mimic the effect of a cosmological constant.  From Eqs. (\ref{maineq}) and (\ref{oscillate}), a present-day
value of $\Omega_\Lambda \approx 0.7$ can be achieved for $A \omega = 5.4 \times 10^{-18}$ s$^{-1}$.
Additionally, we require that
$A \ll 1$ and $\omega t_0 \gg 1$, where $t_0$ is the present-day Hubble time, in order
for Eq. (\ref{Aomega}) to be valid and the first term in Eq. (\ref{maineq}) to be ignored.  While one could,
in principle, choose sufficiently small $A$ and sufficiently large $\omega$ to evade all other observational
constraints on time-varying $G$, it is far from clear that this could originate in any realistic model.

\begin{acknowledgments}

We are grateful to A.-K. Burns, T.M.P. Tait, and M. Valli for providing access to an early verson of
PRyMordial, and for helpful discussions regarding the code.

\end{acknowledgments}

\end{document}